# Stochastic Entanglement Configuration for Constructive Entanglement Topologies in Quantum Machine Learning with Application to Cardiac MRI

Mehri Mehrnia
*Radiology, Biomedical Enginnering*
Northwestern University
Chicago, IL, United States
mehri.mehrnia@northwestern.edu

Mohammed S.M. Elbaz
*Radiology, Biomedical Enginnering*
Northwestern University
Chicago, IL, United States
mohammed.elbaz@northwestern.edu

*Abstract*— Efficient entanglement strategies are critical for advancing variational quantum circuits (VQCs) in hybrid quantum–classical neural networks (QNNs). However, existing approaches predominantly rely on fixed entanglement topologies that are not adaptive to task-specific requirements, limiting their potential to outperform classical models. To address this limitation, we propose a stochastic entanglement configuration method that systematically and adaptively generates diverse entanglement topologies to identify a subspace of constructive entanglement configurations—defined here as configurations that enhance hybrid model performance (e.g., classification accuracy) relative to classical baselines. In our method, each entanglement configuration is represented as a stochastic binary matrix, where entries denote the presence of directed entanglement between qubits. This formulation enables scalable exploration of the entanglement design space using two key metrics: entanglement density (the proportion of entangled qubits) and per-qubit constraints (the number of entanglements initiated per qubit). We define two sampling modes: an unconstrained mode, allowing variable entanglement per qubit, and a constrained mode, which enforces fixed entanglement per qubit. Using the proposed method, we generated 400 stochastic entanglement configurations and applied them to a hybrid QNN for cardiac magnetic resonance imaging (MRI)-based disease classification. Our technique successfully identified 64 (16%) constructive configurations that consistently outperformed the classical baseline. Ensemble aggregation of top-performing configurations achieved a classification accuracy of ~0.92, outperforming the classical model (~0.87) by over 5%. Remarkably, when compared to four conventional entanglement topologies/configurations (ring, nearest-neighbor, no entanglement, and fully entangled), none of these configurations surpassed classical performance (maximum accuracy ~0.82), while our identified configurations delivered up to ~20% higher accuracy performance—highlighting the robustness and generalizability of the newly identified constructive entanglement configurations.

*Keywords*—Stochastic Entanglement Configuration, QNN, Variational Quantum Circuit, Configurable Ansatz, Cardiac MRI disease

I. INTRODUCTION

Quantum computing is an emerging computational paradigm that leverages the principles of quantum mechanics – such as superposition and entanglement – to process information in fundamentally new approaches. Its integration with machine learning known as Quantum Machine Learning (QML) has recently gained increasing interest for its potential to solve complex learning tasks more efficiently than classical approaches [1], [2]. QML models have emerged as a prominent strategy for achieving quantum advantage on near-term, noisy intermediate-scale quantum (NISQ) devices[3], [4].

Among QML architectures, Variational Quantum Circuits (VQCs) are particularly prominent. These VQC circuits employ parameterized quantum gates optimized via gradient-based methods to minimize a cost function, analogous to how classical neural networks are trained [3]. A practical realization of VQCs is through hybrid quantum–classical models, where a pre-trained classical neural network is combined with a trainable quantum circuit to form an end-to-end architecture. Schuld et al. introduced a circuit-centric quantum classifier using amplitude encoding and variational circuits optimized via hybrid training[5]. Hybrid QML models integrating parameterized quantum circuits with classical learning layers are promising tools for tasks requiring high expressivity and low-resource learning. Landman *et al.* demonstrated a quantum neural network(QNN) capable of medical image classification, highlighting the potential of QNNs in real-world imaging applications [6]. Matic et al. demonstrated that QNNs, trained end-to-end with a CNN and a shallow quantum layer using fixed linear entanglement, can classify radiological images. To tackle data scarcity and improve model adaptability, Quantum Transfer Learning (QTL) emerged to offer a promising strategy for medical image analysis. One widely adopted QTL framework is the DressedQuantumNet, which places the quantum circuit between classical pre- and post-processing layers[7]. Variations of this method have been successfully applied to various medical classification tasks such as breast cancer detection, histopathology classification, and COVID-19 diagnosis using chest X-rays[8], [9], [10], [11], [12], [13], [14]. While these QML models and approaches are promising, elegant, and compact, they rely on typical fixed entanglement and did not explore impact of entanglement structure on quantum learning performance. Notably, existing QML models remain typically unable of outperforming classical models counterparts [15].

Despite the growing adoption of VQCs, a critical open question remains: what role does quantum entanglement play in learning performance? Entanglement is often assumed to be essential for enabling quantum advantage. However, several recent studies have raised skepticism about its necessity and practical contribution to learning performance—particularly in small-scale quantum circuits. Some empirical findings suggest that entanglement may not consistently improve solution quality and can even degrade performance [16], [17] [18]. These findings raise important questions about how entanglement should be configured within VQCs to maximize model expressivity and generalization[18]. These insights underscore the importance of ansatz design—not only in terms of entanglement structure but also in dynamic tailoring of entanglement architecture to the specific task or dataset. Problem-specific ansatz design may play a critical role in fully harnessing the expressive power of VQCs while maintaining trainability and robustness. Nevertheless, existing QML models predominantly rely on fixed or hardware-imposed entanglement topologies, such as ring or nearest-neighbor [7], [19], which are not adaptable to the specific requirements of different tasks. The lack of flexible and task-adaptive entanglement strategies remains a key limitation, in part due to the absence of systematic methods for generating or identifying entanglement configurations that can boost quantum learning performance beyond classical counterparts.

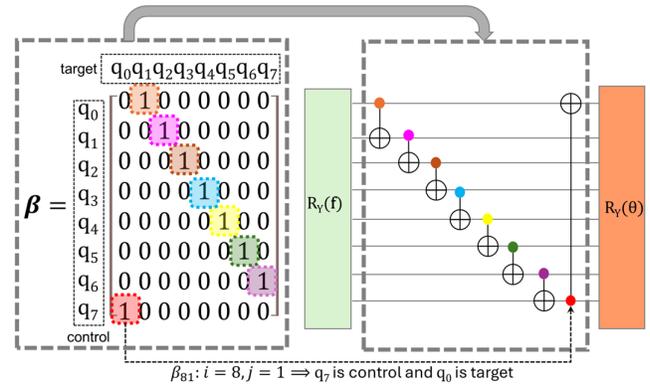

Fig1. Schematic representation of a stochastic entanglement configuration defined by its binary matrix **β**. Each entry $\beta_{ij} = 1$ denotes a CNOT gate with qubit $i$ as the control and qubit $j$ as the target. The matrix shown encodes a well-known entanglement configuration(ring) with $E = 8$ total entanglements, entanglement density $\mu = 16\%$, and a per-qubit constraint $k = 1$ (each row contains a single 1). Each entanglement ($\beta_{ij} = 1$) is color-coded according to its control qubit and mapped to the corresponding quantum circuit on the right. The circuit consists of an initial $R_Y(f)$ layer, followed by the entanglement block defined by **β**, and concludes with an $R_Y(\theta)$ layer.

To address this challenge, we propose the concept of *constructive entanglements*, defined as entanglement configurations or topologies that leverage optimized qubit associations which complement each other, thereby maximizing the expressivity of quantum circuits for specific qubit-encoded features and enabling QML models to outperform their classical counterparts. We hypothesize that a subspace of such constructive entanglements exists within the broader hyperspace of candidate entanglement configurations for a given set of input qubits, enabling optimized quantum learning by maximizing these complementary qubit associations. Here, we present a novel stochastic method for variational quantum circuits that systematically traverses this entanglement hyperspace, generating and evaluating diverse configurations to flexibly identify a subspace of constructive entanglement topologies that significantly boost QML model performance beyond classical baselines.

Unlike conventional quantum models that rely on fixed or hardware-imposed static entanglement patterns (e.g., ring or nearest-neighbor) [7], [19], our method uses stochastic binary matrices to represent flexible, directed entanglement between qubits (Fig. 1). Each matrix element denotes the presence or absence of entanglement between a control and target qubit, allowing arbitrary, task-specific connectivity patterns. To control the space of configurations, we introduce two key design metrics: 1) *Entanglement density*—the total proportion of entangled qubit pairs, and 2) *Per-qubit constraint*—the number of entanglements initiated by each qubit as a control.

These parameters enable exploration of entanglement topologies across a continuum from sparse to dense configurations, and facilitate the identification of patterns that yield robust learning performance.

Given the growing role of machine learning in personalized medicine, QML offers a promising approach for medical applications and new capabilities for tackling high-dimensional and data-limited problems in clinical imaging tasks[20]. Here, we evaluated our method in the context of a clinically relevant medical imaging task: classification of myocardial infarction (MI) using cardiac MRI images, the gold standard for detecting infarcted myocardial (MI) heart disease which represents injury to the heart muscle typically due to heart attack [21], [22]. We employ the task of classifying MRI images as diseased i.e. presenting MI, or normal (no MI). Using the publicly available cardiac MRI dataset with expert labeled ground truth, to evaluate our method, we embed our stochastic entanglement framework into a hybrid quantum–classical model using DressedQuantumNet [7]. The classical front-end employs a pre-trained ResNet18 [23] to extract high-level features from MRI images, followed by dimensionality reduction via PCA[24] and quantum encoding via angle encoding into a low-depth quantum circuit. The quantum layer is then instantiated using our proposed method to sample a diverse set of entanglement configurations, enabling non-linear transformation in a high-dimensional Hilbert space. A final classical layer performs classification (Fig. 2).

Our results show that the proposed method identifies a constructive subspace of 64 new flexible entanglement configurations that consistently improve model performance. Specifically, we demonstrate that:

1)Top-performing configurations outperform classical baselines by over 5%, and

2) Ensembles of top-performing configurations outperform best-performing conventional entanglement topologies by up to ~20% higher accuracy (average 10% above top performing topology), under the same experimental settings.

These findings reinforce the importance of task-adaptive entanglement design and the potential of our proposed stochastic configuration methods to unlock new performance gains in quantum machine learning.

The remainder of this paper is structured as follows. Section II formulates the proposed stochastic entanglement configuration method within the DressedQuantumNet. Section III outlines the experimental setup, including dataset, preprocessing, and training setup. Section IV and V present and discuss the performance results across various entanglement configurations. Finally, Section VI concludes with key findings and directions for future research.

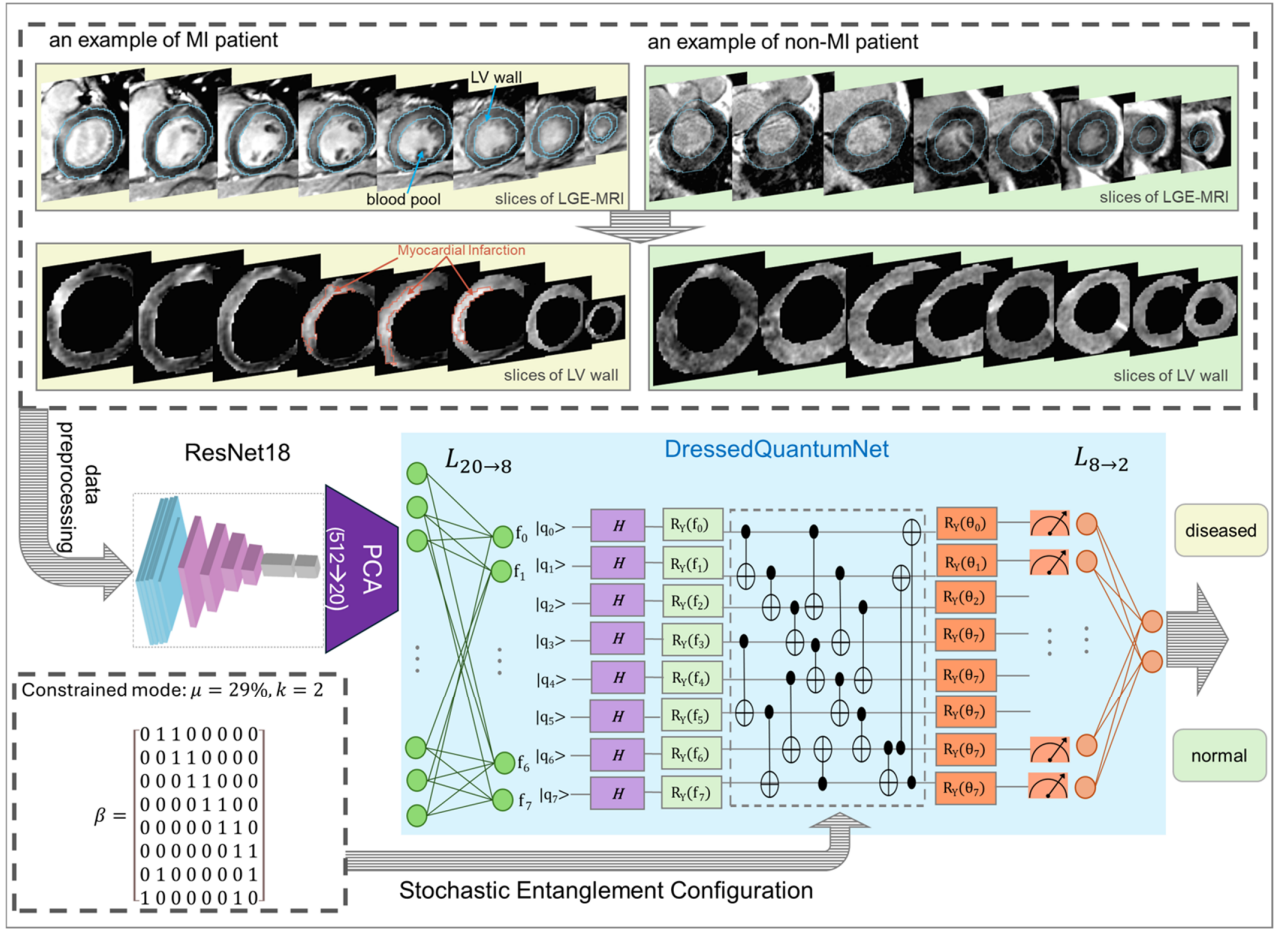

Fig.2. Overview of stochastic entanglement configuration within hybrid quantum-classical pipeline with application for cardiac disease classification from LGE-MRI images of left ventricle (LV). Top: LV myocardial wall is segmented from LGE-MRI using ground truth.
Preprocessed myocardial wall images showing a patient with myocardial infarction (MI) (yellow box) and another patient without MI (green box). The slices of patient without MI are categorized as normal. Bottom: Hybrid-classical quantum model with ResNet18 as feature extractor and PCA for dimension reduction. In the DressedQuantumNet architecture the features are passed through a fully connected layer for further dimension reduction and encoded into a variational quantum circuit using angle encoding. The stochastic entanglement configuration (β matrix) governs circuit entanglements, and quantum measurements are further processed by classical layers for binary classification. Note: The difference in contrast is just for visualization purposes.

## II. METHODS

To address the challenge of identifying effective entanglement topologies in variational quantum circuits (VQCs) in QML, we propose a stochastic formulation of entanglement configuration. Specifically, we represent each entanglement configuration as a random binary matrix, where entries denote the presence of directed entanglement between pairs of qubits. This formulation allows systematic exploration of a high-dimensional space of possible entanglement topologies and enables the identification of constructive (performance-enhancing) entanglement configurations/topologies. We define two distinct stochastic modes: an unconstrained mode, allowing a variable number and distribution of entanglements per qubit, and a constrained mode, in which each qubit is restricted to initiating a fixed number of entanglement connections.

The overall architecture, illustrated in Fig. 2, combines classical and quantum components in a hybrid learning framework. Preprocessed cardiac MRI images are input to a ResNet18 backbone for deep feature extraction, followed by principal component analysis (PCA) for dimensionality reduction. The resulting low-dimensional features are compressed and angle-encoded into parameterized quantum circuits within the DressedQuantumNet architecture, where the entanglement configuration is defined by the stochastic binary matrix.

### A. Stochastic Entanglement Configurations

We define stochastic entanglement configurations using a binary matrix where each array denotes the presence or absence of an entanglement from a control qubit to a target qubit. The binary entanglement matrix $\boldsymbol{\beta} \in \{0,1\}^{n_q \times n_q}$ with $n_q$ denotes the number of qubits in the system as,

$$\boldsymbol{\beta} = \begin{bmatrix} 0 & \cdots & \beta_{1n_q} \\ \vdots & 0 & \vdots \\ \beta_{n_q 1} & \cdots & 0 \end{bmatrix} \quad (1)$$

where $\beta_{ij} \in \{0,1\}$ indicates the presence (1) or absence (0) of an entanglement between control qubit $i$ with target qubit $j$, Fig.1. The following convention is adopted:

- Control qubit corresponds to row index $i$
- Target qubit corresponds to column index $j$

To prevent self-entanglement, we impose the constraint $\beta_{ii} = 0$ for all i = 1, ..., $n_q$, Fig.1.

This formulation supports both symmetric and asymmetric entanglement topologies, depending on the type of entanglement gate employed (e.g., symmetric for CZ and asymmetric for CNOT). For asymmetric entanglement (e.g., directional gates like CNOT), each pair $i,j$ are treated independently. Thus, the number of possible entanglement configurations are:

$$|\mathcal{B}|_{asym} = 2^{n_q(n_q-1)} \quad (2)$$

For symmetric entanglement (e.g., undirected gates like CZ), the matrix is upper triangular such that $\beta_{ij} = \beta_{ji}$. In this case, the number of possible configurations is reduced by half,

$$|\mathcal{B}|_{sym} = 2^{\frac{n_q(n_q-1)}{2}} \quad (3)$$

Due to the exponential growth of entanglement patterns, exhaustive evaluation is infeasible. We adopt a scalable sampling strategy with constraint to explore stochastic entanglement configurations efficiently. We first introduce a metric called entanglement density.

*a) Entanglement density*

First, we define the number of entanglements initiated by specific qubit $i$ as,

$$E_i = \sum_{j=1, j\neq i}^{n_q} \beta_{ij} \quad (4)$$

$E_i$ denotes the number of entanglements initiated by qubit $i$ where $\beta_{ij} = 1$ indicates an entanglement from control qubit $i$ to target qubit $j$. $E_i$ can vary from 0, meaning qubit $i$ initiates no entanglements, to $n_q - 1$, where it is entangled with all other qubits.

Then total number of entanglements, $E$, permitted within a stochastically-defined binary entanglement matrix, $\boldsymbol{\beta}$, is denoted as,

$$E = \sum_{i=1}^{n_q} E_i \quad (5)$$

To normalize this value and quantify the density of the binary entanglement matrix, we define the entanglement density, $\mu$, as the ratio of total number of entanglements, $E$, relative to the maximum possible number of entanglements in which every qubit is entangled with all qubits except itself,

$$\mu = \frac{E}{n_q(n_q-1)} \times 100 \quad (6)$$

The allowable range of the entanglement ratio, $\mu$, spans from 0%—representing no entanglement—to 100%, in which all qubits are entangled.

To enable scalable and interpretable exploration of the exponentially large entanglement configuration space, we introduce a dual-constraint formulation that combines global entanglement density with local per-qubit constraints. To operationalize this formulation, we define two stochastic entanglement configuration modes—a constrained mode and an unconstrained mode—as described next.

*b) Unconstrained Entanglement Mode*

In the unconstrained mode, no restriction is imposed on the number of entanglements each qubit can initiate as a control qubit, $E_i \in \{0,1,..,n_q - 1\}$.

*c) Constrained Entanglement Mode*

In the special case of constrained entanglement mode, we define uniform entanglement configuration in which all qubits initiate exactly $k$ entanglements as control qubits,

$$E_i = k, \forall i \in \{1,..,n_q\} \quad (7)$$

where $k$ is constant and its allowable range is $k \in \{1,..,n_q - 1\}$. This results in a total number of entanglements,

$$E = n_q \times k \quad (8)$$

Such a uniform pattern is particularly advantageous for analyzing the isolated impact of entanglement configuration on model behavior, as it eliminates variability across qubits while maintaining a fixed entanglement density $\mu$ defined as,

$$\mu = \frac{n_q \times k}{n_q(n_q-1)} \times 100 = \frac{k}{n_q - 1} \times 100 \quad (9)$$

The dual-constraint entanglement configuration—consisting of both global entanglement density and local constrained entanglement mode—facilitates the generation of structured, interpretable, and computationally manageable subsets of the exponentially large stochastic entanglement configuration space, enabling systematic evaluation of how different entanglement configuration affect entire architecture performance.

*d) A Special Case o Semi-Constrained Entanglement*

We also consider a semi-constrained configuration mode where each qubit initiates a variable number of entanglements, but the number is bounded within a predefined set,

$$E_i \in \{0,1,..,k_{max}\}; k_{max} \leq n_q - 1 \quad (10)$$

This mode generalizes the constrained mode by allowing heterogeneity across qubits, while still limiting the complexity of the entanglement topology.

*B. DressedQuantumNet*

To facilitate the integration of classical features into quantum circuits, we adopt the DressedQuantumNet architecture [7], a hybrid classical-quantum model in which a variational quantum circuit (VQC) is augmented with two trainable classical layers. These classical layers serve to enhance and streamline the encoding and decoding processes. Specifically, the input features are first dimensionally reduced and non-linearly transformed via a classical layer to match the number of available qubits.

The resulting lower-dimensional feature vector is then mapped to quantum states using a layer of Hadamard gates and angle encoding. The VQC processes these quantum states through trainable single-qubit rotations and entangling gates. The resulting Pauli-Z expectation values are decoded by a final classical layer to produce classification logits. All components are optimized jointly via end-to-end training.

The DressedQuantumNet is denoted as[7],

$$\tilde{Q} = L_{n_q \to n_{out}} \circ Q \circ L_{n_{in} \to n_q} \quad (11)$$

where Q is VQC, $L_{n_1 \to n_2}$ is classical fully connected layer with $n_1$ input and $n_2$ output units as,

$$L_{n_1 \to n_2}(\mathbf{x}) = \varphi(\mathbf{Wx} + \mathbf{b}) \quad (12)$$

Here, $\mathbf{W} \in \mathbb{R}^{n_1 \times n_2}$ is the weight matrix, $\mathbf{b} \in \mathbb{R}^{n_2}$ is the bias vector, and $\varphi$ is a non-linear activation function.

The first classical layer $L_{n_{in} \to n_q}$ compresses and transforms input features into a dimension compatible with the quantum encoder with $n_q$ qubits, while the final layer $L_{n_q \to n_{out}}$ maps Pauli-Z expectation outputs to class logits.

The dimensionally reduced features obtained through transformation by the classical layer, $L_{n_{in} \to n_q}$, are subsequently embedded into quantum states using Hadamard gates and angle encoding. In this process, the classical feature vector $\mathbf{f} = (f_1, \dots, f_{n_q}) \in \mathbb{R}^{n_q}$ is mapped into the Hilbert space, $\mathcal{H} = (\mathbb{C}^2)^{\otimes n_q}$,

$$|\psi(\mathbf{f})\rangle = \otimes_{i=1}^{n_q} R_Y(f_i) H |0\rangle = \cos(f_i)|0\rangle + \sin(f_i)|1\rangle \quad (13)$$

where $H$ is the Hadamard gate and each feature component $f_i$ parametrizes a single-qubit rotation (e.g., $R_Y(f_i)$).

Following the encoding stage, the quantum circuit applies a stochastic entanglement configuration using CNOT gates. Subsequently, quantum features are processed by a $R_Y(\theta_i)$ where each qubit $i$ is rotated by a trainable angle $\theta_i$ around the Y-axis. The quantum parameters, $\boldsymbol{\theta}$, are optimized by the classical component of the hybrid algorithm, thereby closing the quantum–classical training loop. The entanglement pattern, combined with the variational rotations, enables a nonlinear mapping of the input features into a higher-dimensional Hilbert space, allowing the model to learn more expressive and discriminative representations.

Following the $R_Y(\boldsymbol{\theta})$ rotation layer, expectation values of the Pauli-Z operators are measured across all $n_q$ qubits. These measurement outcomes are then passed through the final classical layer of $L_{n_q \to n_{out}}$ to produce the output logits for classification.

### C. Feature extraction and dimensionality reduction

To provide compact and informative inputs to the DressedQuantumNet, we first extract high-level features from each input image using ResNet18[23], pre-trained on ImageNet. The output is a 512-dimensional feature vector, which exceeds the dimensionality supported by DressedQuantumNet. To address this limitation, we apply PCA to reduce the feature dimension, making it suitable for quantum embedding while preserving the most significant information from the input.

The ResNet18's final classification layer is removed, and the output of its penultimate layer serves as a 512-dimensional feature vector,

$$\mathbf{x} = \text{ResNet18}(\mathbf{I}), \mathbf{x} \in \mathbb{R}^{512} \quad (14)$$

where $\mathbf{I} \in \mathbb{R}^{3 \times H \times W}$ represents the 3-channel input image. Since the feature dimensionality exceeds the input size supported by DressedQuantumNet, PCA [24] is applied to obtain a projection matrix $\mathbf{W}_{PCA} \in \mathbb{R}^{512 \times d}$, which is then used to compute the dimensionally reduced feature vector,

$$\mathbf{z} = \mathbf{x} \mathbf{W}_{PCA}, \mathbf{z} \in \mathbb{R}^d \quad (15)$$

### D. Dataset of cardiac MRI for myocardial infarction disease classification

For method evaluation, we used the public dataset EMIDEC[25], which includes 100 cardiac MRI (Late Gadolinium Enhancement, LGE[26]) scans—67 scans from patients diagnosed with myocardial infarction (MI) and 33 from non-MI patients. In total, the dataset comprises 702 images from all patients. Each scan consists of a series of 2D short-axis images covering the left ventricular (LV) of heart, accompanied by expert-annotated segmentations of the myocardial wall, blood pool and the infarcted (diseased) regions(see examples of MI/non-MI in Fig.2). Importantly, the dataset provides expert classification of images as diseased images i.e. containing MI or a normal (non-MI) that served as the ground truth classification herein. The task here is to classify cardiac MRI images to one of two classes: diseased (i.e. with MI) or normal (i.e. without MI).

## III. EXPERIMENTS

### A. Hybrid Quantum-Classical Implementation

We evaluated the application of stochastic entanglement configurations in a cardiac MRI disease classification task i.e. classifying images to diseased or normal. Specifically, we assessed their influence on the performance of our hybrid classical–quantum model, DressedQuantumNet, which employs 8 qubits, $n_q = 8$. Each stochastic entanglement configuration was represented as an 8×8 binary matrix, where entries indicate the presence or absence of directed entanglement between qubits.

PCA was computed on the training set and applied consistently across the validation and test sets, resulting in a projection matrix $\mathbf{W}_{PCA} \in \mathbb{R}^{512 \times d}$ with $d = 20$.

The classical components of the DressedQuantumNet consist of a fully connected input layer from PCA with $n_{in} = 20$ and an output layer with $n_{out} = 2$ corresponding to the diseased and normal classes.

### B. Classical Baseline Model for comparison

To compare our method with the purely classical model, we implemented a baseline by removing the variational quantum circuit (VQC) from DressedQuantumNet resulting in solely of two fully connected layers of $L_{n_q \to n_{out}} \circ L_{n_{in} \to n_q}$.

### C. Stochastic Entanglement Configurations Experiments

We trained the hybrid model of Fig.2 using 8×8 binary entanglement matrices of $\beta$, evaluating three entanglement densities: $\mu = 14\%$ (E = 8), $\mu = 29\%$ (E =16), and $\mu = 43\%$(E = 24). Each density was explored under:

- **Unconstrained mode**: No restrictions on the number of entanglements per qubit.

- **Constrained mode**: Each entanglement density $\mu = 14\%, 29\%, 43\%$ corresponds to a fixed per-qubit constraint of $k = 1,2,3$, respectively, ensuring that each qubit initiates exactly $k$ entanglements as a control qubit.

For each entanglement density and mode, 50 independent runs were performed, totaling 300 runs of stochastic entanglement configurations. Additionally, a special case of 100 runs of stochastic entanglement configurations were conducted where qubits participated in a bounded variable number of entanglements $E_i \in \{0,1,2,3\}; i = 1,...,8; k_{max} = 3$ which resulted in a variable entanglement density, $\mu$, of [0%,43%]. In total, 400 independent runs of stochastic entanglement configurations were conducted.

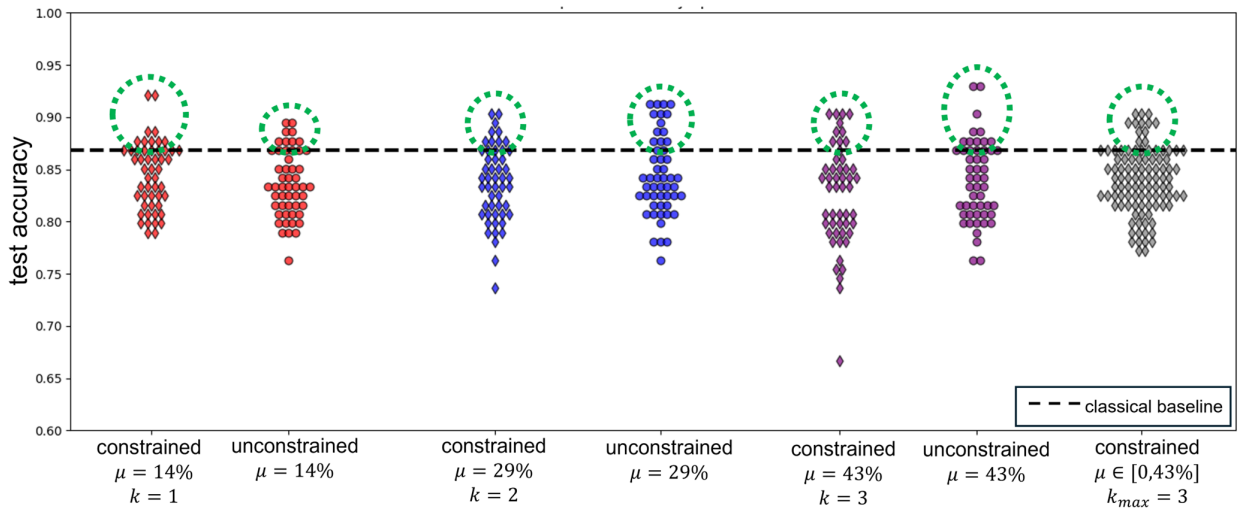

Fig.3. Test accuracies across stochastic entanglement configurations with varying entanglement densities and per-qubit constraints. Each color pair of scatter plots represents a fixed entanglement density: 14% (red), 28% (blue), and 43% (purple). Within each pair, diamond markers correspond to per-qubit constrained entanglement, while circle markers represent unconstrained configurations. The final group on the far right (gray) corresponds to configurations with variable entanglement density, $\mu \in [0,43\%]$, and per-qubit constraint of $k_{max} = 3$, covering all previous stochastic configurations in a more generalized form. The dashed horizontal black line indicates the classical test accuracy (0.8684). Several constructive entanglement topologies—encircled with dashed green ovals—demonstrate superior performance, exceeding the classical baseline across different entanglement densities and w/o constraint.

The hyperparameters used in our experiments were optimized using Optuna[27]. Specifically, the model was trained for 70 epochs using a learning rate of 0.00043, a learning rate scheduler with decay factor (gamma) of 0.6, and a batch size of 8. Our implementation utilizes the PennyLane library integrated with PyTorch.

### D. Comparison to Conventional Entanglement Topologies

To benchmark the performance of newly-derived entanglement configurations by our proposed stochastic entanglement configuration method, we compared the performance to four well-established conventional entanglement topologies using the same hybrid quantum–classical pipeline explained in the above section. These included: ring entangled, nearest-neighbor entangled, no entanglement, and fully entangled topological configurations [7], [19]. Importantly, these widely-used topologies can all be reproduced as special cases within our stochastic method by appropriately selecting the entanglement density, $\mu$, and the per-qubit entanglement constraint, $k$, in Eqns. 6-10 above.

These topologies can be generated as follows:

- *Ring topology* can be generated using $\mu \approx 14\%$ and $k = 1$ connecting each qubit to its immediate neighbor in a circular layout.

- *Nearest-neighbor topology* arises with $\mu \approx 12\%$ and $k = 1$, connecting linearly adjacent qubits.

- *No entanglement topology* is realized by setting $\mu = 0\%$ and $k = 0$, such that no qubit-to-qubit entanglement connections are introduced.

- *Fully entangled topology* corresponds to $\mu = 100\%$ with $k = n_q - 1$ (where $n_q$ is the number of qubits), resulting in all-to-all entanglement.

By enabling the generation of these conventional configurations through stochastic parameterization, our method provides a unified framework for comparing both conventional and adaptive entanglement designs. This also facilitates controlled evaluations across a spectrum of entanglement complexities, ensuring methodological consistency while supporting the discovery of task-specific quantum circuit topologies.

### E. Data Preparation and Preprocessing

For the purpose of cardiac MRI disease classification task, we assigned labels to each image as *diseased* or *normal*, based on a presence of ground-truth MI region. For non-MI patients, all images were labeled as normal. For MI cases, labeling was performed on a per-slice basis using ground-truth MI segmentation masks to determine the presence of the disease.

To ensure robust training and evaluation, the dataset was divided in a patient-wise manner to prevent data leakage across subsets. Specifically, 226 slices (approximately 50%) from 56 patients were used for training, 112 slices (~25%) from 22 patients for validation, and 114 slices (~25%) from 22 patients for testing. All slices belonging to a given patient were assigned exclusively to a single subset to maintain subject-level independence.

The region of interest (ROI) in this study is the myocardial wall (Fig. 1). To ensure that feature learning is focused on clinically relevant structures, we apply the provided ground-truth segmentation masks to isolate the myocardial region and exclude background tissue. To further minimize the influence of non-informative background pixels, the masked background is filled with intensities from the myocardial wall, reducing the contrast between foreground and background as well as enhancing model focus on relevant anatomical features.

To match the input requirements of ResNet18, each LGE-MRI 2D image is resized to 224×224 pixels using bilinear interpolation and applying intensity normalization. As ResNet18 expects RGB input, each grayscale image is replicated across three channels to form a 3-channel image. The original LGE-MRI volumes, provided in NIfTI (.nii) format, were converted into 2D images and saved in NumPy (.npy) format. This preprocessing step facilitated efficient data loading and seamless integration with our PyTorch-based training pipeline.

## IV. RESULTS

### A. Performance Analysis of Stochastic Entanglement Configurations versus Classical Baseline

The results of all 400 runs of the hybrid quantum–classical model, each with a distinct stochastic entanglement configuration, are visualized in Fig.3, showing test accuracy across different entanglement densities and constraint modes. Several configurations surpass the classical baseline test accuracy of 0.8684 (indicated by the horizontal dashed black line), with high-performing configurations marked by dashed green ovals. In total, 64 configurations (16%) outperformed the classical baseline, underscoring the ability of strategically sampled entanglement patterns to enhance cardiac disease classification performance and identify constructive subspace of entanglements. These high-performing configurations were observed across both constrained and unconstrained modes, with test accuracy ranging from 0.8772 to 0.9298.

TABLE I. TEST ACCURACY OF STOCHASTIC ENTANGLEMENT CONFIGURATIONS USING MAJORITY VOTING ALONG WITH CONVENTIONAL ENTANGLEMENT CONFIGURATIONS

| Stochastic entanglement configuration metrics | Test Accuracy | Number of stochastic configurations |
|---|---|---|
| *Test Accuracy using Ensemble Configuration Majority Voting* | | |
| Constrained $\mu = 14\%, k = 1$ | 0.8596 | 50 |
| Unconstrained $\mu = 14\%$ | **0.8860** | 50 |
| Constrained $\mu = 29\%, k = 2$ | **0.8772** | 50 |
| Unconstrained $\mu = 29\%$ | **0.8860** | 50 |
| Constrained $\mu = 43\%, k = 3$ | **0.8772** | 50 |
| Unconstrained $\mu = 43\%$ | 0.8684 | 50 |
| Constrained $\mu \in [0,43\%], k_{max} = 3$ | **0.8947** | 100 |
| *Test Accuracies by Top-r% Ensemble Configuration Majority Voting* | | |
| Unconstrained $\mu = 29\%$ | **0.9211** | Top - 5 % of runs |
| *Test Accuracies using Conventional Entanglement Configurations* | | |
| Ring topology $\mu = 14\%, k = 1$ | 0.8246 | 1 |
| nearest neighbor topology $\mu = 12\%$ | 0.8158 | 1 |
| no entanglement $\mu = 0$ | 0.8246 | 1 |
| fully entangled $\mu = 100\%$ | 0.7368 | 1 |

Particularly, the constrained stochastic entanglement configurations with entanglement densities of $\mu = 14\%, 29\%, 43\%$ with per-qubit constraint of $k = 1,2,3$, achieved the top test accuracies of 0.9211, 0.9035, and 0.9035, respectively. In the unconstrained mode, these values are 0.8947, 0.9123, and 0.9298, respectively. All outperforming the accuracy of classical baseline counterpart, 0.8684.

We also visualize the learning trajectories of quantum parameters(**θ**) across different entanglement configurations with both high and low performance. As shown in Fig.4.a, the configuration with an entanglement density of $\mu = 29\%$ exhibits dynamic and diverse parameter evolution throughout training and reached to test accuracy of 0.9123. In contrast, lower-performing configurations—such as those with $\mu = 20\%$ and $\mu = 18\%$, both yielding a test accuracy of 0.7719—often exhibit early saturation or stagnation across multiple parameters, Fig.4.d,e. For example, qubits 2 and 3 in Fig. 4.d and qubits 4 and 6 in Fig. 4.e appear isolated. This behavior suggests limited expressivity in circuits with low entanglement density particularly where certain qubits remain unentangled and isolated from the broader quantum interaction.

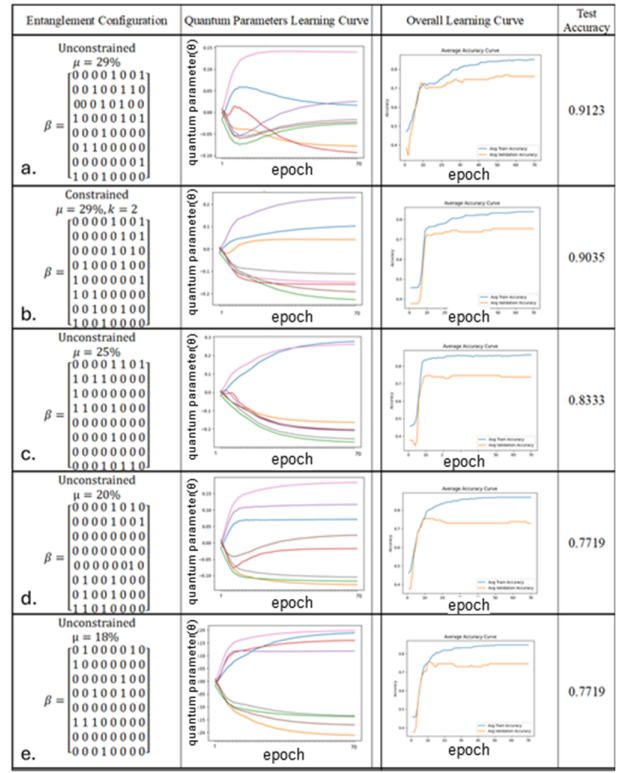

Fig.4. Examples of training dynamics across stochastic entanglement configurations with varying entanglement densities ($\mu$) and per-qubit constraints ($k$). Each row corresponds to a distinct entanglement configuration. The first column shows the binary matrix of entanglement configuration, **β**. The second column illustrates the evolution of quantum parameters during training, **θ**. The third column presents the corresponding average training and validation accuracy curves over 70 epochs. The rightmost column reports the final isolated test accuracy achieved for each configuration.

### B. Performance Comparison to Conventional Entanglement Topologies

We further compared our methods' newly derived entanglement results to set of four well-established entanglement topology configurations. These included: no entanglement, fully entangled, ring, and nearest-neighbor topologies. Our derived stochastic entanglement configurations outperformed the tested conventional topologies by up to 20% in classification accuracy. The fully entangled configuration—corresponding to an entanglement density of $\mu = 100\%$ and a constrained mode of $k = 7$ — resulted in a lower accuracy of 0.7368 (~20% below ours). The nearest-neighbor configuration ($\mu = 12\%$), implemented with linear qubit adjacency, resulted in an accuracy of 0.8158 (~11% below ours). The ring topology ($\mu = 14\%, k = 1$), in which each qubit connects to its immediate neighbor in a circular manner, achieved test accuracy of 0.8246 (~10% below ours). Finally, The no entanglement configuration ($\mu = 0\%$) also yielded a test accuracy of 0.8246 (~10% below ours)., Table I.

This suggests that over-entanglement could reduce the model's generalizability and its ability to learn discriminative patterns by diluting meaningful interactions between qubits. Excessive entanglement may interfere with qubits' ability to influence each other effectively or result in counteracting and suppressing the localized qubit associations essential for productive quantum learning.

### C. Performance Analysis of Ensemble Majority Voting versus Classical Baseline

To further evaluate the robustness and generalization of stochastic entanglement configurations across various entanglement densities and modes (constrained and unconstrained), we employed a probability-based majority[28] voting strategy on the test set. For each configuration, final class predictions were derived by aggregating SoftMax probabilities from independently trained models. The class with the highest aggregated probability was selected as the final prediction for each test sample.

Using majority voting, we evaluated the aggregated ensemble performance of the top 1%, 5%, 10%, 20%, and 30% of stochastic entanglement configurations, ranked by validation accuracy across predefined entanglement densities in both constrained and unconstrained modes. The corresponding test accuracies are presented in Fig. 5, illustrating not only peak performance but also the consistency of high-performing configurations.

Multiple newly identified constructive configurations surpassed the classical baseline test accuracy of 0.8684, indicated by the bold black dashed line. This underscores the robustness and generalizability of the optimized entanglement topologies.

The highest test accuracy of 0.9211 was obtained by selecting the top - 5% (i.e., 2 out of 50 configurations) from the set with an entanglement density $\mu = 29\%$— representing a performance gain of ~5% over the classical baseline of 0.8684. This was followed by the top - 1% (i.e., the best-performing model out of 50 stochastic entanglement configurations) from the set with entanglement density of $\mu = 43\%$, and per-qubit constrained $k = 3$, achieved a test accuracy of 0.9035- exceeding the classical baseline by over 3%.

## V. DISCUSSION

Our results demonstrate that the topology of entanglement configurations within variational quantum circuits (VQCs) plays a critical role in shaping quantum machine learning (QML) performance. Notably, we confirm the existence of a *constructive entanglement subspace*—a subset of configurations that leverage optimized qubit associations or connections which complement each other, thereby enhancing the expressivity of the quantum circuit for specific qubit-encoded features. By flexibly identifying and employing entanglement topologies that maximize these complementary associations for a given input set of qubit-encoded features, constructive entanglements enable quantum models to more effectively capture and learn complex data patterns, leading to consistent outperformance of classical models under comparable conditions. Utilizing our novel stochastic entanglement configuration technique, we systematically generated and evaluated 400 stochastic entanglement topologies, successfully identifying 64 new constructive configurations that consistently improved classification accuracy performance, surpassing the classical baseline by more than 5% and up to 20% (average 10% over top performing) over conventional entanglement topologies.

Notably, configurations with moderate entanglement density of (e.g., $\mu = 29\%$) consistently yielded strong test performance, especially in constrained mode (i.e., $k = 2$). Sparse circuits (e.g., $\mu = 18\%, 20\%$) underperformed were often associated with underutilized qubits (i.e., inactive qubits which are not involved in entanglement), as reflected by early saturation of quantum parameters during training. Conversely, denser topologies ($\mu = 43\%$) provided good performance in select cases but introduced increased variability across configurations. However, the fully entangled configuration— corresponding to an entanglement density of $\mu = 100\%$ and a constrained mode of $k = 7$ —resulted in significantly lower accuracy. This suggests that balanced entanglement, rather than minimal or excessive connectivity, may offer an optimal trade-off between expressivity and trainability.

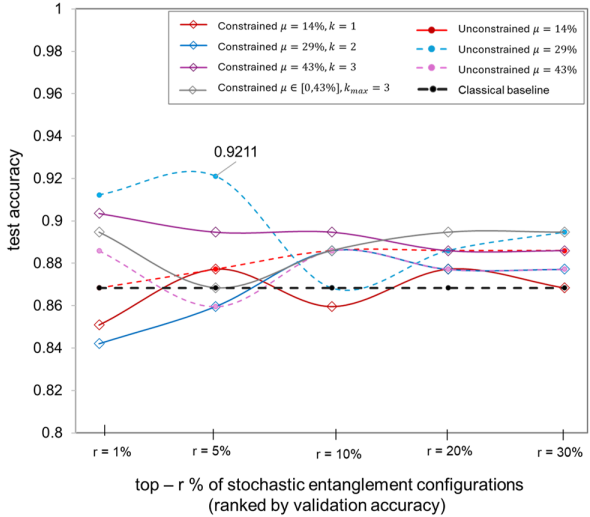

Fig. 5. Majority voting-based test accuracy across top-r% ranked stochastic entanglement configurations, stratified by validation performance. Test accuracy is evaluated for the top 1%, 5%, 10%, 20%, and 30% of configurations across different entanglement densities ($k$) and per-qubit constraint modes ($k$). Solid lines represent configurations with fixed per-qubit entanglement constraints ($k = 1, 2, 3$), while dashed lines correspond to unconstrained configurations. The bold black dashed horizontal line indicates the classical baseline with test accuracy of 0.8684.

The use of majority voting ensembles, constructed from the best-performing configurations selected based on validation performance, yielded the highest accuracy on the isolated test dataset. Aggregating predictions across top-performing configurations not only improved test set accuracy beyond the classical baseline, but also highlighted the consistency of high-performing configurations within the entanglement configuration metrics. These results indicates the robustness and generalizability of the identified entanglement configurations by our proposed method. Notably, the best-performing quantum ensemble ($\mu = 29\%$, unconstrained) reached a test accuracy of 0.9211, outperforming the classical baseline (accuracy 0.8684) by over 5%, Fig.5. This improvement is substantial, especially considering the limited circuit depth of one in our architecture and limited number of 8 qubits, reinforcing the potential of entanglement as a powerful architectural tool.

Remarkably, the new configurations identified by our method outperformed all tested conventional entanglement topologies—including ring, nearest-neighbor, no entanglement, and fully entangled topologies—by more than 10%. Notably, under the same experimental conditions, none of these standard configurations were able to outperform the classical model, underscoring their limited task adaptability and further validating the need for data-driven, flexible entanglement design.

Our technique also uncovered an important distinction: entanglement can be either constructive or destructive. Out of the 400 tested configurations, only 64 (16%) were identified as constructive (identified per validation dataset), consistently improving test performance in isolated dataset beyond the classical baseline model. The remaining 336 (84%) were destructive, yielding no improvement or even worse performance than classical baseline. This experimental observation provides a methodological explanation for conflicting reports in the QML literature regarding efficacy of entanglement [16], [17]. Our findings suggest that while entanglement can be beneficial, the vast majority of configurations are ineffective. Achieving a quantum advantage therefore depends on the careful selection or design of constructive entanglement patterns, underscoring the need for reliable methods to identify such subspaces.

Our proposed technique provides a systematic methodology for identifying these constructive entanglement subspaces. The fact that dozens of effective configurations were identified, rather than a single isolated solution, demonstrates the generality and robustness of our approach. Although validated here in a cardiac MRI classification setting, the technique is domain-agnostic and can be applied to diverse learning tasks.

Interestingly, we observed that different entanglement matrices can yield identical performance, suggesting a level of expressive equivalence in the entanglement space. This raises intriguing theoretical questions about the connectivity-expressivity trade-off in quantum models and points toward the existence of multiple entanglement paths leading to the same decision boundaries. Further analytical studies on qubit connectivity especially in Pennylane simulation platform may help formalize this observation.

While the results are promising, our study is not without limitations. The stochastic nature of the configuration space, combined with computational constraints (running each entanglement configuration requires up to one hour), limited the total number of configurations we were able to evaluate. It is possible that additional high-performing, or even optimal, configurations exist within the unexplored entanglement space. However, our primary aim was to test the hypothesis that a flexible and constructive entanglement subspace exists. Our results confirm this hypothesis and introduce a systematic approach for identifying such subspaces. This also opens new avenues for developing more optimized approaches to traverse large configuration subspaces in future work.

Future work will investigate optimized guided sampling techniques, such as reinforcement learning or Bayesian optimization, to enable more efficient navigation of this high-dimensional space. To further evaluate the generalizability and practical utility of our approach, future work will explore its application across a wider spectrum of datasets, modalities, and network architectures in diverse problem settings[29], [30], [31], [32]. In this work, we initialized quantum parameters randomly. However, prior studies showed that structured initializations, such as Gaussian or layer-wise strategies, can help avoiding barren plateaus by preserving gradient flow—an avenue worth exploring in future research [33], [34]. Additionally, we plan to explore a larger number of qubits to accommodate higher-dimensional features with reduced compression, deeper variational ansätze and adaptive entanglement strategies tailored to task-specific learning dynamics. A valuable next step would be to evaluate the models on quantum hardware to assess their behavior and stability under realistic noise conditions. The use of controlled quantum gates will be also explored to enable more structured and adaptable entanglement configurations, enhancing the quantum circuit's capacity to capture task-specific complexity[35].

## VI. Conclusion

Here, we proposed a novel stochastic entanglement configuration technique that allows efficient and flexible traversing of the high-dimensional space of possible entanglement topologies to identify a subspace of new constructive entanglement topology configurations for variational quantum circuits in QML. Applied to cardiac MRI classification, our method effectively identified a subspace of 64 constructive entanglements, consistently outperforming classical baselines. Ensemble aggregation of the identified top-performing entanglement configurations results in outperforming classical baselines by over 5% in classification accuracy, and surpassed conventional entanglement topologies by up to ~20% in classification accuracy confirming the practical advantage of our method. Future research will focus on higher number of qubits, deeper variational ansatz, guided sampling strategies, adaptive entanglement configurations, different datasets and modalities to further improve performance, generalizability and applicability across diverse real-world tasks.

### Acknowledgment

We acknowledge funding from the National Institutes of Health (NIH), National Heart, Lung, and Blood Institute (NHLBI), grant number R01HL169780.